\documentclass[preprint,amsmath,aps,nofootinbib]{revtex4}
\usepackage{graphicx}

\def\bfsigma{\mbox{\boldmath $\sigma$}}

\def\bfepsilon{\mbox{\boldmath $\epsilon$}}

\def\OMIT#1{}

\newcommand{\nn}{\nonumber}

\newcommand{\bqa}{\begin{eqnarray}}
\newcommand{\eqa}{\end{eqnarray}}

\begin{document}



\title{Exclusive $\eta_b$ decay to double $J/\psi$ at next-to-leading order in $\alpha_s$}

\author{Bin Gong\footnote{E-mail: twain@ihep.ac.cn}}
\affiliation{Institute of High Energy Physics, Chinese Academy of
Sciences, Beijing 100049, China\vspace{0.2cm}}
\affiliation{Theoretical Physics Center for Science Facilities,
Chinese Academy of Sciences, Beijing 100049, China\vspace{0.2cm}}

\author{Yu Jia\footnote{E-mail: jiay@ihep.ac.cn}}
\affiliation{Institute of High Energy Physics, Chinese Academy of
Sciences, Beijing 100049, China\vspace{0.2cm}}
\affiliation{Theoretical Physics Center for Science Facilities,
Chinese Academy of Sciences, Beijing 100049, China\vspace{0.2cm}}

\author{Jian-Xiong Wang \footnote{E-mail: jxwang@ihep.ac.cn}}
\affiliation{Institute of High Energy Physics, Chinese Academy of
Sciences, Beijing 100049, China\vspace{0.2cm}}
\affiliation{Theoretical Physics Center for Science Facilities,
Chinese Academy of Sciences, Beijing 100049, China\vspace{0.2cm}}

\date{\today\\ \vspace{1cm} }



\begin{abstract}

Within the nonrelativistic QCD (NRQCD) factorization framework, we
calculate the exclusive decay process $\eta_b \to J/\psi J/\psi$ to
next-to-leading order in the strong coupling constant, while at
leading order in the charm quark relative velocity. It is found that
this new contribution to the amplitude is comparable in magnitude
with the previously calculated relativistic correction piece,  but
differs by a phase about $90^\circ$. Including this new contribution
will increase the previous prediction to $\mathcal{B}(\eta_b \to
J/\psi J/\psi)$ substantially, thus brightening the discovery
potential of this clean hadronic decay channel of $\eta_b$ in the
forthcoming LHC experiment.

\end{abstract}

\maketitle

\newpage

The quest for the $\eta_b$ meson, the lowest-lying pseudoscalar
bottomonium state,  started shortly after the discovery of
$\Upsilon$ in 1977.   A thorough understanding of its properties,
such as its mass and decay pattern, will be great benefit to
strengthening our knowledge towards heavy quarkonium
physics~\cite{Brambilla:2004wf}.

Tremendous efforts have been  invested in various high energy
collision experiments during the past thirty years to seek this
elusive particle,  but convincing evidence about its existence has
not emerged until very recently.  After many futile searches, there
finally comes out an exciting news from \textsc{Babar} collaboration
last month~\cite{Aubert:2008vj}, which reported the first
unambiguous sighting of $\eta_b$ with a $10\,\sigma$ significance,
through the hindered magnetic dipole transition process
$\Upsilon(3S)\to \eta_b\gamma$.  It is interesting to note the
rather precisely measured hyperfine $\Upsilon(1S)$-$\eta_b$ mass
splitting, $71.4^{+2.3}_{-3.1}({\rm stat})\pm 2.7({\rm syst})$ MeV,
can play a decisive role in discriminating different quarkonium
models~\cite{Brambilla:2004wf}.

Aside from its mass,  almost nothing is known about the dynamical
properties of the $\eta_b$, {\it e.g.}, its decay pattern. Owning to
the weaker QCD coupling at bottom mass scale,  together with the
copious phase space opened up for innumerable decay channels, one in
general expects that each $\eta_b$ exclusive decay mode may only be
allotted a rather small branching ratio. Indeed, a rough estimate of
some 2-body and 3-body hadronic decay channels of $\eta_b$ supports
this expectation, and probably $\eta_b$ disintegrates primarily into
final states of high multiplicities~\cite{Jia:2006rx}.

Some ``golden" modes have been proposed for hunting $\eta_b$, such
as $\eta_b\to J/\psi J/\psi$~\cite{Braaten:2000cm} and $\eta_b \to
J/\psi\gamma$~\cite{Hao:2006nf,Gao:2007fv}.  Despite very clean
signature thanks to the presence of $J/\psi$,  these decay modes
unfortunately have rather suppressed branching ratios.  Obviously,
the $e^+e^-$ collision experiments like $B$ factory are not well
suited to pin down these decay modes due to the relatively low
$\eta_b$ yield. By contrast, high energy hadron collision
experiments in general possess a much larger $\eta_b$ production
rate,  which in turn allows for triggering these rare but very clean
decay modes.  For the decay $\eta_b\to J/\psi J/\psi$, the original
hope is that it should be discoverable in Tevatron Run
2~\cite{Braaten:2000cm}. Subsequent investigation reveals the
situation may not be so
optimistic~\cite{Maltoni:2004hv,Jia:2006rx}~\footnote{Note there
recently appeared an estimate for this process based on a
final-state interaction model, in which $\eta_b$ first annihilates
into $D \overline{D}^*$, and the charm meson pair subsequently
scatter into double $J/\psi$~\cite{Santorelli:2007xg}. The
corresponding prediction critically hinges on the branching ratio of
$\eta_b\to D \overline{D}^*$,  which is poorly known at present.
However, there is reasonable argument that suggests this ratio may
be rather suppressed~\cite{Jia:2006rx}.}. In particular, an explicit
NRQCD calculation predicts ${\cal B}[\eta_b\to J/\psi J/\psi]\sim
10^{-8}$, which implies that notwithstanding the pessimistic
prospect of observing this decay channel in Tevatron Run 2,  the
chance may still be bright in the forthcoming LHC
program~\cite{Jia:2006rx}.

The exclusive decay $\eta_b\to J/\psi J/\psi$,  apart from its
phenomenological interest,  is worthy of theoretical exploration
in its own right.  Since this hard exclusive process comprises
entirely heavy quarkonium,  it can serve as an ideal laboratory to
test the applicability of NRQCD and perturbative
QCD~\cite{Bodwin:1994jh}. Note this process is quite similar to
the double charmonium production processes at $B$ factory, {\it
e.g.}  $e^+e^-\to J/\psi\eta_c, \, J/\psi J/\psi$, which have
stirred quite a few attention since the first \textsc{Belle}
measurements were released in 2002~\cite{Abe:2002rb} (for an
incomplete list about NRQCD-based investigation,
see~\cite{Braaten:2002fi,Liu:2002wq,Bodwin:2002fk,Bodwin:2002kk,
Hagiwara:2003cw,Zhang:2005ch,He:2007te,Bodwin:2007ga,Gong:2007db,Gong:2008ce}).
No doubt a systematic study of this hadronic decay process will
offer complementary insight towards understanding the decay and
production mechanisms of quarkonium.

A peculiar feature of the decay $\eta_b\to J/\psi J/\psi$ is
that~\footnote{This decay can also proceed through $\eta_b\to
\gamma^*\gamma^* \to J/\psi J/\psi$. Since this QED contribution
is much less important than its QCD counterpart, it will be
neglected in this work.}, at the lowest order in $\alpha_s$ and
charm quark relative velocity in $J/\psi$, $v_c$, the amplitude
{\it vanishes}. Therefore, to obtain a non-vanishing result, one
must proceed to the higher orders either in $\alpha_s$ or $v_c$.
In Ref.~\cite{Jia:2006rx}, the leading relativistic correction to
the tree-level process was explored. In this work, we make
progress along the alternative route, {\it i.e.}, we compute the
${\cal O}(\alpha_s)$ correction but stay at the zeroth order in
$v_c$. Since $\alpha_s\sim v_c^2$, the latter contribution is
expected to be as important as the former one, so it will be
definitely desirable to work out this missing piece.

Before presenting our main result, it is instructive to recall
some dynamical properties about this process.  Let $\lambda$ and
$\tilde{\lambda}$ stand for the helicities of two $J/\psi$ mesons
in the $\eta_b$ rest frame.  According to the helicity selection
rule and angular momentum conservation, the decay configuration
($\lambda$, $\tilde{\lambda}$)=(0, 0)  exhibits the slowest
asymptotic decrease with $m_b$,  ${\cal B}\sim
1/m_b^4$~\cite{Brodsky:1981kj}. However, it is impossible for
$\eta_b$ to decay into two longitudinally polarized $J/\psi$
because of conflict between parity and Lorentz invariance. The
hadron helicity conservation, $|\lambda+\tilde{\lambda}|=0$, can
be violated either by the nonzero charm mass $m_c$ or by the
transverse momentum of $c$ inside $J/\psi$, $q_\perp$.   For every
unit of deviation of this rule, there is a further suppression
factor of $m_c^2/m_b^2$ or $q_\perp^2/m_b^2$.  For the only
physically allowed helicity configurations ($\lambda$,
$\tilde{\lambda}$)=($\pm 1$, $\pm 1$), the selection rule is
violated by two units, so one expects ${\cal B}\sim 1/m_b^8$ no
matter the cause of violation is due to nonzero $q_\perp$ or
nonzero $m_c$.

A noteworthy finding is that, at the lowest order in $\alpha_s$,
the violation of the rule should be ascribable to the nonzero
transverse momentum of $c$ in $J/\psi$,  instead of the nonzero
$m_c$~\cite{Jia:2006rx}.  The corresponding branching ratio
exhibits the following asymptotic scaling
\bqa {\cal B}_{v_c^2} [\eta_b\to J/\psi(\pm 1) +J/\psi(\pm 1)]
&\sim& \alpha_s^2 \,v_c^6 \left(m_c\over m_b\right)^{4} \cdot
\left(q_\perp^2\over m_b^2\right)^{2} \sim \alpha_s^2 \,v_c^{10}
\left(m_c\over m_b\right)^8\,, \label{Br:scaling:relativ:corr} \eqa
where the subscript emphasizes the corresponding contribution comes
from the relativistic correction. The factor $v_c^6$ affiliated with
the would-be ``leading-power" scaling, stems from the squares of the
wave functions at the origin of two $J/\psi$, since
$\psi_{J/\psi}(0)\propto (m_c v_c)^{3/2}$. In the last equation
$q_\perp\sim m_c v_c$ has been assumed.

\begin{figure}[tb]
\begin{center}
\includegraphics[scale=0.7]{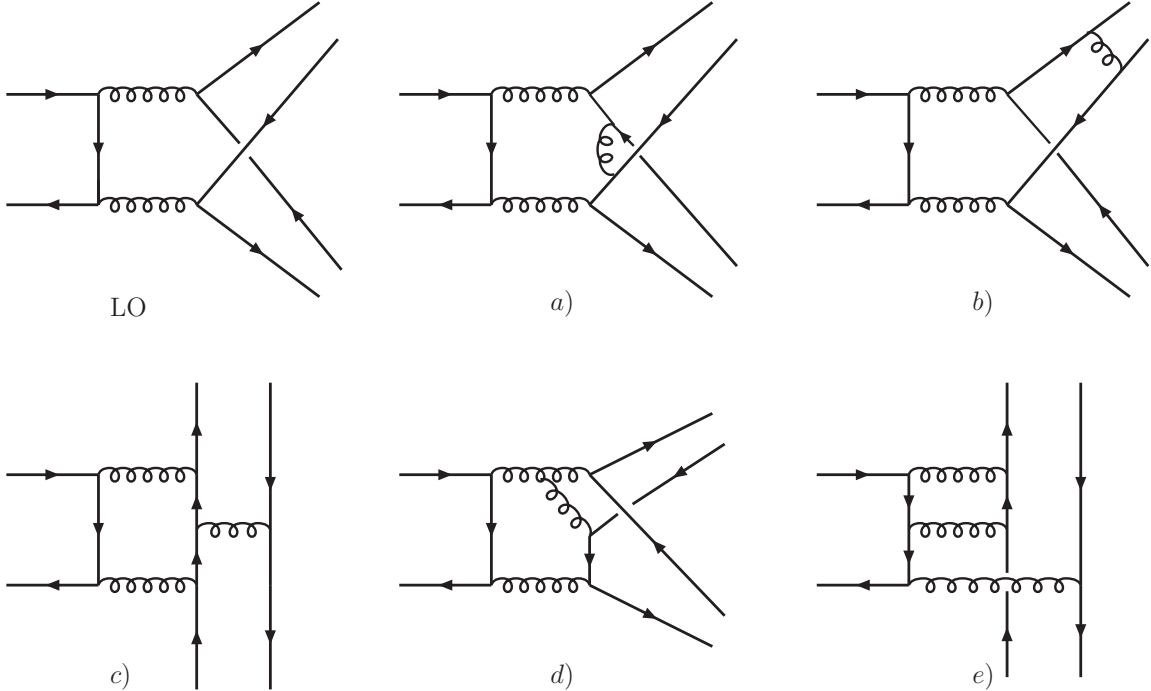}
\caption{Lowest order diagram, together with some representative
next-to-leading order diagrams [from $a)$ to $e)$] that contribute
to $\eta_b \to J/\psi\,J/\psi$. \label{feynman:diag}}
\end{center}
\end{figure}

One may naturally wonder how the situation becomes when one goes to
the next-to-leading order (NLO) in $\alpha_s$.  Some typical NLO
Feynman diagrams contributing to $\eta_b\to J/\psi J/\psi$ are shown
in Fig.~\ref{feynman:diag}.  Should it also be obligatory to retain
finite relative momentum of $c$ in order to obtain a non-vanishing
result?   To answer this question, it is helpful to draw some clue
from an analogous hadronic decay process, $\Upsilon\to
J/\psi\eta_c$, which also violates the helicity selection rule. In
this $\Upsilon$ decay process,  the lowest-order diagrams start
already at one-loop, {\it i.e.}, at ${\cal O}(\alpha_s^6)$, which in
fact share the same topology as those represented  by
Fig.~\ref{feynman:diag}e). There it is found that keeping a nonzero
$m_c$ but neglecting the relative momentum of $c$ is sufficient to
give rise to a non-vanishing result~\cite{Jia:2007hy}.  This may be
reckoned as a persuasive hint that a similar pattern may occur in
our case.  A more direct evidence comes from inspecting the cut
structure of Fig.~\ref{feynman:diag}.  It is  easy to observe that
all the diagrams in Fig.~\ref{feynman:diag} [except
Fig.~\ref{feynman:diag}e)] permit a specific cut that has a clear
physical meaning,  {\it i.e.},  a cut through two intermediate gluon
lines that divides the full process into $\eta_b\to gg$ and $gg\to
J/\psi J/\psi$.  Since both sub-processes are finite at the zeroth
order in $v_c$,  so does the corresponding portion of the imaginary
part of the $\eta_b\to J/\psi J/\psi$ amplitude which is obtained by
stitching both on-shell sub-amplitudes together in accordance with
the recipe of Cutkosky rule.  Since a fraction of the imaginary part
of the total amplitude is nonzero, there is no reason to expect the
full imaginary part should vanish, thus we are reassured that at
this order, the nonzero $m_c$ can adequately play the role of
violating the helicity selection rule, and it suffices to set the
relative velocity $v_c$ to zero in the calculation.  As a result, we
expect the NLO perturbative contribution scales as follows
\bqa {\cal B}_{\alpha_s} [\eta_b\to J/\psi(\pm 1) +J/\psi(\pm 1)]
&\sim& \alpha_s^2 \,v_c^6 \left(m_c\over m_b\right)^{4} \cdot
\,\alpha_s^2\left(m_c^2\over m_b^2\right)^{2} \sim \alpha_s^4
\,v_c^6 \left(m_c\over m_b\right)^8\,, \label{Br:scaling:radia:corr}
\eqa
where the subscript stresses the corresponding contribution comes
from the ${\cal O}(\alpha_s)$ perturbative correction.  As will be
confirmed by the explicit calculation, this power-law scaling indeed
holds besides logarithmical scaling violation effect.

Having said this much,  let us now proceed to the actual
calculation. We first set up some notations.   Let
$P,\varepsilon\, (\tilde{P},\tilde{\varepsilon})$ signify the
momentum and polarization vector of each $J/\psi$. Owning to
parity and Lorentz invariance, the decay amplitude ${\cal M}$ is
constrained of the following structure:
\bqa
{\cal M} &=& {\cal A} \: \epsilon^{\mu \nu \alpha \beta}\,P_\mu \,
\tilde{P}_\nu \,\varepsilon^*_\alpha(\lambda) \,
\tilde{\varepsilon}^*_\beta(\tilde{\lambda})\,.
\label{Lorentz:tensor:structure}
\eqa
Apparently,  the only allowed helicity configurations indeed are
$(\lambda,\tilde{\lambda})=(\pm 1,\pm 1)$. All the nontrivial
dynamics is encoded in the reduced amplitude ${\cal A}$, and our
task is then to unravel its explicit form.

In total 80 diagrams can be drawn for this process at ${\cal
O}(\alpha_s^6)$.  A full treatment of these diagrams, at first
sight, may seem to be a daunting task. Luckily, the calculation
turns out to be much simpler than one would imagine,  thanks to some
gratifying traits of this process.  First, renormalization is not
needed in this work.  This originates from the fact that the
amplitude at leading order in $\alpha_s$ and $v_c$ vanishes. Since
the contribution of the counterterm diagrams are proportional to the
LO diagram, they simply vanish too.  As a consequence, all the
diagrams which contain primitive ultraviolet (UV) divergent subgraph
must instead be free of the UV divergences,  otherwise these
unremoved divergences would signal the failure of the
renormalization theory.  A remarkable property of this process is
that, these superficially UV-divergent diagrams are not only finite,
in fact all of them {\it vanish}. Similarly, among the remaining
manifestly UV-finite diagrams, a simplification further occurs,
namely many diagrams with a symmetric topology also {\it vanish},
{\it e.g.}, diagrams involving one 4-gluon vertex, diagrams
involving two 3-gluon vertices, and diagrams containing a gluon
ladder between the $b$ and $\bar{b}$ lines.  This can be mainly
attributed to the specific Lorentz structure as indicated in
(\ref{Lorentz:tensor:structure}), that is, in these diagrams, the
Levi-Civita tensor that arises from the $b\bar{b}$ annihilation
amplitude,  is either contracted with some symmetric Lorentz
indices,  or there are no sufficient number of independent 4-vectors
to contract with it.

There finally survive 40 diagrams with non-vanishing contribution,
some of which are shown in Fig.~\ref{feynman:diag}a)-e).  In fact,
because of the interchanging symmetry between two $J/\psi$, diagrams
of a given topology usually give rise to an identical answer, and
practically only much fewer diagrams need to be calculated.  We
employ the automated Feynman Diagram Calculation package (FDC) to
fulfill the analytic evaluation of the required one-loop diagrams.
FDC is a powerful program based on the symbolic language
\textsc{Reduce},  which is designed to automate the perturbative
quantum field theory calculation in computer. FDC was initially
developed by one of us long ago~\cite{Wang:1993rt}, and the function
of automatic one-loop calculation has recently been realized by two
of us~\cite{Gong:2008:thesis}.  Recently, FDC has been successfully
applied to several important quarkonium production processes to
compute the NLO perturbative
corrections~\cite{Gong:2007db,Gong:2008ce}.

In passing, we would like to comment on some specific issues in
the calculation. One technical complication is  the occurrences of
six-point one-loop integrals in Fig.~\ref{feynman:diag}a) and b).
FDC has implemented some systematic recursive algorithm to reduce
higher-point integrals down to lower-point ones. In this work, it
turns out that all the encountered 6-point integrals can be
reduced to the 4-point integrals and lower ones. In addition,
Passarino-Veltman algorithm~\cite{Passarino:1978jh} has also been
built into FDC to expedite the reduction of tensor integrals to
the scalar ones.

It turns out that each individual diagram is infrared finite,  at
least in the Feynman gauge as we worked with.  At a glance, one may
feel that some care needs to be paid in handling
Fig.~\ref{feynman:diag}b), since the Coulomb divergences may
potentially arise there.  Hearteningly, a close examination shows
that this diagram is in fact free of any singular $1/v_c$ poles. The
absence of Coulomb singularity can be most transparently understood
in the NRQCD factorization ansatz~\cite{Bodwin:1994jh}. What we can
calculate within perturbation theory is the so-called {\it hard}
(short-distance) coefficient.   All the long-distance singularities
encountered in computing the one-loop QCD diagrams, must be exactly
reproduced in the respective one-loop NRQCD diagrams with a {\it
potential} gluon attached between the outgoing $c$-$\bar{c}$ lines
that forms a $J/\psi$,  so they ultimately cancel out in the
intended hard coefficient via the matching procedure. However, the
corresponding diagrams in the NRQCD sector trivially vanish in our
case,  again because the $\eta_b\to J/\psi J/\psi$ process vanishes
at LO in $\alpha_s$ and $v_c$,  therefore no any net long-distance
divergence, such as Coulomb divergence, should be expected to appear
in the diagrams in the QCD sector.

Finally it may be worthwhile to compare our process with
$\Upsilon\to J/\psi\eta_c$.  As stated earlier, the latter process
is described by a small subset of Fig.~\ref{feynman:diag}, the class
of diagrams exemplified by Fig.~\ref{feynman:diag}e),  in which
three ``Abelian" gluons connect between $b$ quark line and the charm
quark lines. Since $\Upsilon$ has negative $C$ parity, charge
conjugation invariance demands that three gluons must form a totally
symmetric color-singlet state, therefore only the fully symmetric
part of the color factor $d^{abc}d^{abc}$ can contribute in this
$\Upsilon$ decay process~\cite{Jia:2007hy}.  On the contrary, due to
the positive $C$ parity of $\eta_b$,  in our case the three gluons
must instead be totally antisymmetric in color space, so the
$f^{abc}f^{abc}$ piece finally survive in the color factor.  We have
explicitly checked this is indeed the case.

\begin{figure}[tb]
\begin{center}
\includegraphics[scale=0.6]{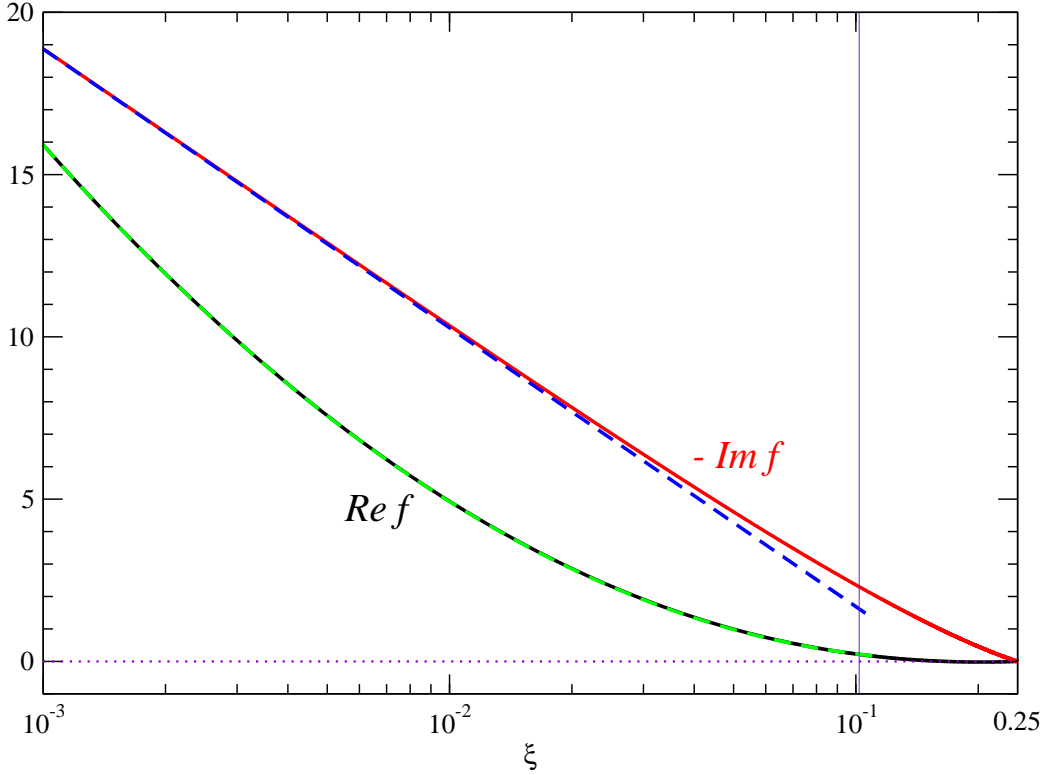}
\caption{The solid lines represent the real and imaginary parts of
$f(\xi)$,  the dashed lines represent their respective asymptotic
expressions as given in (\ref{Re:f:asymptotic}) and
(\ref{Im:f:asymptotic}).  The vertical line marks the place
$\xi=m_c^2/m_b^2=0.102$, which is adopted in the phenomenological
analysis. \label{plot:f}}
\end{center}
\end{figure}

After analytic manipulation of all the diagrams by FDC, together
with some simplifying by hand,  we can express the reduced amplitude
as following:
\bqa
{\cal A}_{\alpha_s} &=& {512\sqrt{6}\, \pi  \,\alpha_s^3 \,m_c
\over 27 \,m_b^{9/2} {\bf P}^2} \psi_{\eta_b}(0) \,
\psi_{J/\psi}^2(0) \,f\left({m_c^2}\over m_b^2\right)\,,
\label{red:am:NLO:alphas}
\eqa
where the relativistic normalization is used for the quarkonium
states, and $|{\bf P}|=\sqrt{m_b^2-4m_c^2}$ is the magnitude of
$J/\psi$ momentum. The complex-valued $f$ function, which has
encompassed the loop contribution, is deliberately chosen as
dimensionless, therefore it can depend on $m_b$ and $m_c$ only
through their dimensionless ratio $m_c^2/m_b^2$. Furthermore, the
$f$ function is normalized such that it varies with $m_c^2/m_b^2$
only logarithmically, which is slower than any power law scaling.
After some quick algebra, one can verify that
Eq.~(\ref{red:am:NLO:alphas}) is in conformity to ${\cal
B}_{\alpha_s} \sim 1/m_b^8$ up to logarithmical modification.

We display the shape of the $f$ function in Figure~\ref{plot:f}. The
analytic expression for the real part of the $f$ function is too
lengthy to be reproduced here. However, the analytical result for
the imaginary part is quite compact,
\bqa {\rm Im}f(\xi)& = & {\pi \over 8} \left\{ {1\over \xi
(2+\xi)}-{\beta\over 2\xi} -{63+86\xi \over 2(2+\xi)}+{2\over
\beta}\left[-\left({23\over 2}-14\xi\right) \tanh^{-1} \beta
\right.\right.
\nn \\
&+& \left.\left. (53-14\xi)\tanh^{-1} \left({\beta\over 3}\right)
-(1-2\xi)\ln(2-4\xi)-(1+2\xi)\ln(4\xi)\right]
  \right\}\,,
\label{Im:f} \eqa
where $\beta=\sqrt{1-4\xi}$.

It is instructive to deduce the asymptotic behavior of $f$ in the
small $\xi$ limit. After some efforts, we obtain
\bqa {\rm Re}f_{\rm asym}(\xi)&=& {19\over 32}\ln^2 \xi +
\left({15\over 16}+ {13\over 8 }\ln2 \right)\ln \xi -{53\over 16}
\ln^2 2 + {45\over 8}\ln 2-{31\over 32}
\nn \\
&-& {5 \pi^2\over 32} + {3\,\sqrt{3}\over 8}\pi + {\cal
O}(\xi\ln^2\xi) \,,\label{Re:f:asymptotic}
\\
{\rm Im}f_{\rm asym}(\xi)& = & \pi \left({19\over 16}\,\ln \xi +
{47\over 8}\ln 2 -{15\over 8}+ {\cal O}(\xi\ln\xi)
\right)\,.\label{Im:f:asymptotic} \eqa
As can be clearly visualized in Fig.~\ref{plot:f}, as $\xi \to 0 $,
the real and imaginary parts of the $f$ function exhibit $\ln^2 \xi$
and $\ln \xi$ dependence, respectively. This seems to be a generic
trait of exclusive bottomonium decays to double charmonium,  which
is also seen in the analogous $f$ function in the $\Upsilon\to
J/\psi\eta_c$ process~\cite{Jia:2007hy}. It is interesting to note
that, up to the physically relevant point $\xi=m_c^2/m_b^2= 0.102$,
which is obtained by substituting $m_c =1.5$ GeV and $m_b=4.7$ GeV,
the asymptotic expressions in (\ref{Re:f:asymptotic}) and
(\ref{Im:f:asymptotic}) still constitute decent approximations to
the true expressions, where the agreement for the real part is
especially satisfactory, and the agreement for the imaginary part is
at 70\% level.  This may suggest that,  if one is willing to carry
out this calculation by hand yet without loss of accuracy, the most
efficient way is to first expand the integrand in $m_c^2/m_b^2$
prior to performing the loop integrals.

It is interesting to note that, provided that $\xi$ is not overly
small, say, $\xi>3.8 \times 10^{-4}$, then $-{\rm Im}\,f$ is always
bigger than $|{\rm Re}\, f|$. In particular, at the
phenomenologically relevant point $\xi=0.102$,  we find the
corresponding $f= 0.216-2.30 i=2.31 \, e^{-i \, 84.6^\circ}$,  and
the real part seems to be practically negligible.

Since the NLO perturbative amplitude is dominated by its imaginary
part,  one may like to speculate which type of cuts might constitute
the most important contribution. It might sound attractive to
inquire whether that specific cut as mentioned earlier, the one
acting upon two intermediate gluon lines that partitions the full
process into $\eta_b\to gg$ and $gg\to J/\psi J/\psi$,  would
dominate over all other possible cuts.  If this were the case, we
would be happy because this process then could be endowed with a
``plausible" physical picture -- that it proceeds first through
$\eta_b$ annihilation into two gluons, and these two gluons
subsequently rescatter into double $J/\psi$.   Let us examine this
conjecture concretely.  With the aid of the Cutkosky rule, it is
straightforward to deduce this particular fraction of the imaginary
part of the $f$ function, which we call ${\rm Im}\,f_{2g}$:
\bqa {\rm Im}\,f_{2g}(\xi)& = & -\pi \left\{ 2+{1\over
16\xi}-{9\over 4 \beta} \tanh^{-1}\beta \right\}.
\label{Im:f2g:cut} \eqa
Note the occurrence of the  $1/\xi$ term,  which can be traced to
the cut contribution from Fig.~\ref{feynman:diag}a), renders this
function scale with quark masses as $m_b^2/m_c^2$,   so that the
true scaling ${\cal B}_{\alpha_s} \sim 1/m_b^8$ is violated.  This
is unacceptable and clearly indicates approximating the full
imaginary part by this specific fraction is unjustified. Numerically
for $\xi=0.1$, we obtain ${\rm Im}\,f_{2g}=1.15$, which is far off
the true value ${\rm Im}\,f=-2.30$. A careful investigation reveals
that each diagram in Fig.~\ref{feynman:diag} [except
Fig.~\ref{feynman:diag}e)] possesses a rich cut structure, for
example,  Fig.~\ref{feynman:diag}a), b), d)  allow for 6, 5, and 4
different kind of cuts, respectively.  Among all the possible cuts,
those which pass through one gluon line and another nonadjacent
quark line generally lead to more involved expressions in
(\ref{Im:f}). For a given diagram, it seems not profitable to single
out one specific cut to mimic its full imaginary part. For example,
only after the remaining cuts are included, the imaginary part of
Fig.~\ref{feynman:diag}a) then has a sensible answer, free of the
dangerous $1/\xi$ singularity.

We are now very close to the intended formula. For completeness, we
still need the expression of the reduced amplitude that arises from
the tree-level relativistic correction~\cite{Jia:2006rx}
\bqa {\cal A}_{v_c^2} &=& {512 \sqrt{6}\, \pi^2 \,\alpha_s^2 \,m_c
\over 27 \,m_b^{13/2} } \,\psi_{\eta_b}(0) \, \psi_{J/\psi}^2(0)
\,\langle v^2\rangle_{J/\psi}\,. \label{red:am:rel:corr} \eqa
Roughly speaking,  $\langle v^2 \rangle_{J/\psi}$ is a quantity
related to the second derivative of the wave function at the origin
for $J/\psi$, which governs the size of relativistic corrections. In
fact, this factor admits a rigorous definition as a ratio of NRQCD
matrix elements~\cite{Bodwin:1994jh,Braaten:2002fi}:
\bqa \langle v^2 \rangle_{J/\psi} &=& {\bfepsilon \cdot \langle
 J/\psi(\bfepsilon)|\psi^\dagger \bfsigma (-{\bf D}^2) \chi |0 \rangle \over m_c^2 \,
\bfepsilon \cdot \langle J/\psi(\bfepsilon)|\psi^\dagger \bfsigma
\chi |0 \rangle }, \eqa
where $\bfepsilon$ denotes the polarization vector of $J/\psi$ at
its rest frame,  $\psi$ and $\chi$ represent Pauli spinor fields
in NRQCD,  $\mathbf D$ is the spatial covariant derivative.  The
matrix elements appearing in the above ratio are understood to be
the renormalized ones. It is worth stressing that the sign of
$\langle v^2 \rangle_{J/\psi}$ needs not necessarily to be
positive.

Plugging equations (\ref{red:am:NLO:alphas}) and
(\ref{red:am:rel:corr}) in the formula
\bqa \Gamma[\eta_b \to J/\psi J/\psi] &=& {|{\bf P}|^3 \over
8\pi}\,|{\cal A}_{\alpha_s}+{\cal A}_{v_c^2}|^2\,, \eqa
we then obtain the desired partial width.   Note the cubic power of
momentum reflects that two $J/\psi$ are in relative $P$-wave state.
This formula has already taken into account the sum over transverse
polarizations of two $J/\psi$  as well as their
indistinguishability. It may be more convenient to present an
explicit expression for the branching ratio, where the wave function
at the origin of $\eta_b$ drops out,
\bqa {\cal B}[\eta_b\to J/\psi\,J/\psi]&=&  K_{gg}^{-1} \, {2^{13}
\,\pi^2 \,\alpha_s^2 \over 3^4 } \,{m_c^2 \over m_b^8}
\,\psi^4_{J/\psi}(0)\,\left(1-{4 m_c^2\over m_b^2}\right)^{3/2}
\nn \\
&\times & \left|\langle v^2 \rangle_{J/\psi}\, + {\alpha_s \over
\pi} \left(1-{4 m_c^2\over m_b^2}\right)^{-1} f\left({m_c^2}\over
m_b^2\right) \right|^2\,. \label{BR:Etab:psipsi} \eqa
In deriving this, we have approximated the total width of $\eta_b$
by its hadronic width:
\bqa \Gamma_{\rm tot}[\eta_b] &\approx & \Gamma[\eta_b\to {\rm
light\; hadrons}] = K_{gg}\,{8 \pi \alpha_s^2 \over 3 m_b^2}
\,\psi_{\eta_b}^2(0)\,, \eqa
where the factor $K_{gg}$ encodes the magnitude of NLO perturbative
correction to $\eta_b\to gg$~\cite{Petrelli:1997ge}
$$K_{gg}= 1+
\left({53\over 2}-{31 \pi^2\over 24}-{8\, n_f \over
9}\right){\alpha_s(2m_b)\over \pi}.
$$
$n_f$ denotes the number of active light flavors. One usually takes
$n_f=4$, by treating charm quark also as a massless parton. For
$\alpha_s(2m_b)=0.18$, we get $K_{gg}=1.58$.

Equation (\ref{BR:Etab:psipsi}) constitutes the key formula of this
work. From this equation  one can readily confirm the asymptotic
behaviors anticipated in (\ref{Br:scaling:relativ:corr}) and
(\ref{Br:scaling:radia:corr}).

\begin{figure}[tb]
\begin{center}
\includegraphics[scale=0.62]{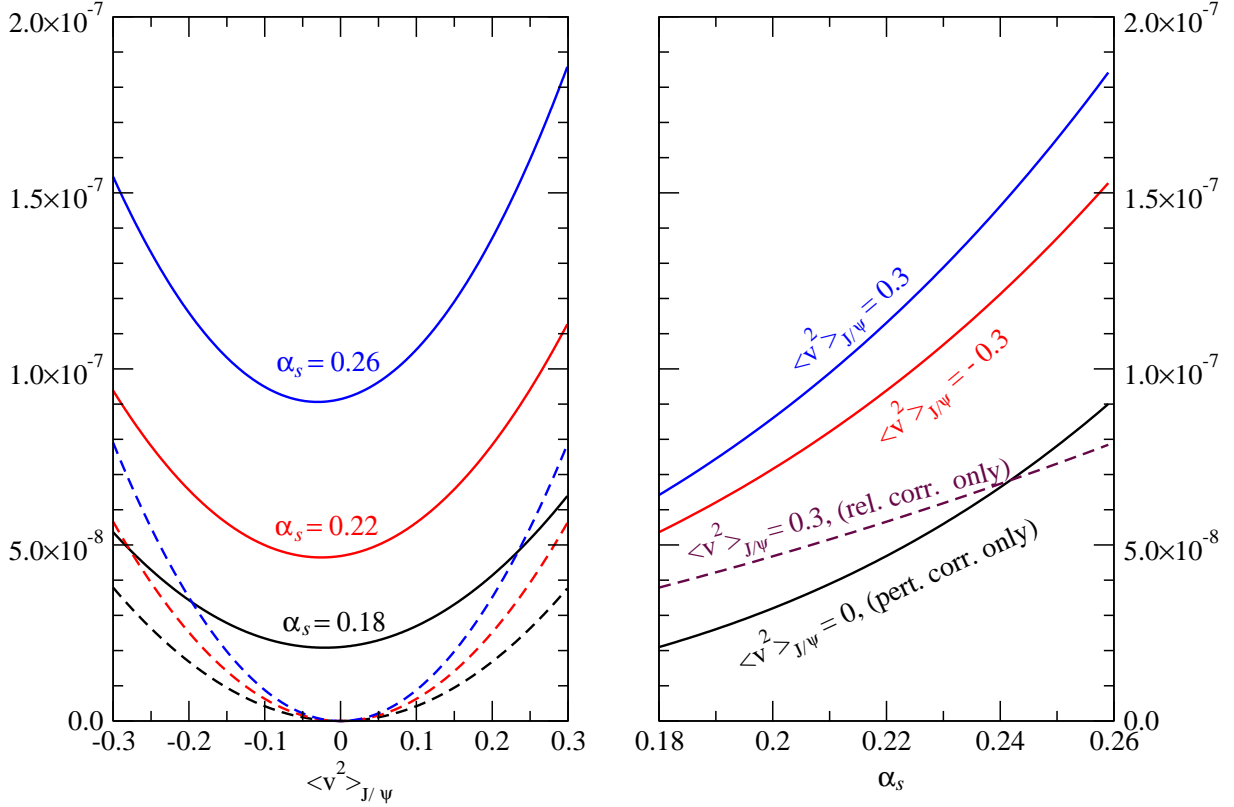}
\caption{${\cal B}[\eta_b\to J/\psi\,J/\psi]$ as a function of
$\langle v^2 \rangle_{J/\psi}$ (left panel) and $\alpha_s$ (right
panel). The solid lines represent the full prediction from
(\ref{BR:Etab:psipsi}) and the dashed lines are obtained by keeping
only the relativistic correction piece. In the left panel, three
different pairs of curves (each pair tinted by the same color) are
obtained by fixing the renormalization scale $\mu $ to be $2m_b$,
$m_b$ and $2m_c$ respectively, so that the corresponding strong
coupling constant $\alpha_s$ is 0.18, 0.22 and 0.26.
\label{plot:br:2:column}}
\end{center}
\end{figure}

Let us now explore the numerical outcome of (\ref{BR:Etab:psipsi}).
The input parameters are $m_b$, $m_c$, $\alpha_s$,
$\psi_{J/\psi}(0)$ and $\langle v^2\rangle_{J/\psi}$, respectively.
We take $m_b =M_{\eta_b}/2 \approx 4.7$ GeV and $m_c=1.5$ GeV.
Obviously, our predictions are quite sensitive to the value assigned
for the strong coupling constant. To account for the affiliated
ambiguity, we slide the renormalization scale $\mu$ from $2m_b$ to
$2m_c$, corresponding to varying $\alpha_s$ between 0.18 and 0.26.
The wave function at the origin of $J/\psi$ can be extracted from
its electric width:
\bqa \Gamma[J/\psi\to e^+e^-]&=& {4\pi e_c^2\alpha^2 \over m_c^2}
\psi^2_{J/\psi}(0)\left(1-{8\,\alpha_s(2m_c)\over 3\pi}\right)^2\,,
\label{jpsi:ee:width:nlo} \eqa
where the first-order perturbative correction has been included.
Using the measured electric width 5.55 keV~\cite{Yao:2006px}, we
obtain $\psi_{J/\psi}(0)=0.263\;{\rm GeV}^{3/2}$.

Among various input parameters, the least known about is $\langle
v^2\rangle_{J/\psi}$.  Nevertheless, there is a useful relation,
first derived by Gremm and Kapustin by employing the equation of
motion of NRQCD~\cite{Gremm:1997dq},  which expresses this quantity
in terms of the charm quark pole mass and the $J/\psi$ mass.  To our
purpose this relation reads~\cite{Braaten:2002fi}
\bqa \langle v^2\rangle_{J/\psi} & \approx & {M_{J/\psi}^2-4\,
m^2_{c\, {\rm pole}}\over 4\, m^2_{c\,{\rm pole}}}\,.
\label{GK:relation} \eqa
Owing to the intrinsic ambiguity associated with the charm quark
pole mass, a wide spectrum of estimates for $m_{c\,{\rm pole}}$ is
scattered in literature.  Ref.~\cite{Braaten:2002fi} takes
$m_{c\,{\rm pole}}=1.4$ GeV in the analysis of double charmonium
production at $B$ factory,  which leads to $\langle
v^2\rangle_{J/\psi}=0.22$ from (\ref{GK:relation}). Note this value
is compatible with $\langle
v^2\rangle_{J/\psi}=0.225^{+0.106}_{-0.088}$, which is obtained from
a recent Cornell potential-model based
analysis~\cite{Bodwin:2006dn,Bodwin:2007fz}.  A positive $\langle
v^2\rangle_{J/\psi}$ is helpful to alleviate the discrepancy between
the LO NRQCD prediction and the measured $J/\psi+\eta_c$ production
rate at \textsc{Belle}.  Notwithstanding the phenomenological
inclination, however, one may be alert that a rather different value
might not be excluded on theoretical ground. For example,  if
$m_{c\,{\rm pole}}=1.75 \pm 0.15$ GeV is assumed, which stems from a
recent QCD moment sum rule analysis~\cite{Eidemuller:2002wk},  one
would instead obtain $\langle v^2 \rangle_{J/\psi}=-0.22\pm 0.15$,
whose sign has even changed.  Because of the impossibility to nail
down $m_{c\,{\rm pole}}$ to an accuracy better than $\Lambda_{\rm
QCD}$, it seems safe to take a conservative attitude, by assigning a
rather large uncertainty to $\langle v^2 \rangle_{J/\psi}$. In this
work, we will assume $-0.3<\langle v^2 \rangle_{J/\psi}<0.3$,
including the possibility that it may even vanish.

In Figure~\ref{plot:br:2:column} we illustrate how the branching
ratio of $\eta_b\to J/\psi\,J/\psi$  varies with $\langle v^2
\rangle_{J/\psi}$  and $\alpha_s$.  In the left panel of
Fig.~\ref{plot:br:2:column}, one can clearly see that inclusion of
the QCD perturbative correction will significantly enhance the
previous prediction that only implements relativistic correction,
especially when $|\langle v^2 \rangle_{J/\psi}|$ is small.   The
approximately parabolic dependence of ${\cal B}$ on $\langle v^2
\rangle_{J/\psi}$, reflects the fact that the NLO perturbative
amplitude, ${\cal A}_{\alpha_s}$,  is dominated by its imaginary
part,  therefore it is nearly orthogonal to the relativistic
correction amplitude ${\cal A}_{v_c^2}$, so their contributions can
be added almost incoherently.  To see the interference pattern
lucidly, we make a numerical presentation to the terms inside the
absolute value sign in (\ref{BR:Etab:psipsi}),
$$ \langle v^2 \rangle_{J/\psi}+ 1.24 \, e^{-i\,84.6^\circ}\alpha_s =
(\langle v^2 \rangle_{J/\psi}+0.12\alpha_s) - i\,1.24 \,\alpha_s.$$
It is interesting to note, besides the approximate $90^\circ$ phase
difference, the (imaginary part of) NLO perturbative contribution is
comparable in magnitude with the relativistic correction piece. The
residual interference between the real part of the perturbative
correction amplitude  and the relativistic correction piece has
modest effect, which can be either constructive or destructive
depending on the sign of $\langle v^2\rangle_{J/\psi}$. In the right
panel of Fig.~\ref{plot:br:2:column}, one can further observe that,
despite the acute sensitivity to the choice of the renormalization
scale, the complete result is considerably larger than each
individual contribution in the considered range of $\alpha_s$.

As can be inferred from Figure~\ref{plot:br:2:column}, our final
estimate of the branching ratio is
\bqa {\cal B}[\eta_b\to J/\psi\, J/\psi] &= & (2.1-18.6)\times
10^{-8}\,. \label{etab:2Jpsi:prediction:new}
\eqa
Not surprisingly,  the uncertainty is rather large.  It is
enlightening to compare our result with the previous prediction that
includes only the relativistic correction
contribution~\cite{Jia:2006rx}:
\bqa {\cal B}_{v_c^2}[\eta_b\to J/\psi\, J/\psi] &= &
(0.5-6.6)\times 10^{-8}\,. \label{etab:2Jpsi:prediction:old} \eqa
Though the strategies of estimating the error differ in both work,
it is nevertheless unambiguous to see the substantial effects of the
NLO perturbative correction. It is worth mentioning that the lowest
bound of our prediction is already as large as $2\times 10^{-8}$,
which can be entirely attributed to the NLO perturbative
contribution. In contrast, the prediction including only the
relativistic correction simply vanishes as $\langle
v^2\rangle_{J/\psi}$ approaches zero.

Finally we present an updated estimate about the discovery potential
of this decay mode in the LHC experiments. In the hadron collision
experiment $J/\psi$ can be cleanly reconstructed via its muonic
decay mode.  Multiplying (\ref{etab:2Jpsi:prediction:new}) by the
branching ratios of 6\% for each of the decay $J/\psi\to \mu^+
\mu^-$, we obtain $ {\cal B}[\eta_b\to J/\psi J/\psi \to 4 \mu]
\approx (0.7-6.7)\times 10^{-10}$.
 Assuming the $\eta_b$ production cross section  at LHC to
 be about 15 ${\rm \mu
b}$~\cite{Jia:2006rx}, one then finds the cross section for the 4
$\mu$ events to be about 1-10 fb.  For a 300 ${\rm fb}^{-1}$ data,
which is expected to be gleaned in one year run at LHC design
luminosity, the number of produced events may reach 300-3000. The
product of acceptance and efficiency for detecting $J/\psi$ decay to
$\mu^+\mu^-$ is estimated to be $\epsilon\approx
0.1$\cite{Braaten:2000cm}, which is perhaps a conservative estimate
for LHC.  Multiplying the number of the produced events by
$\epsilon^2$, we expect between 3 and 30 observed events per year.
From this rough estimate, we are tempted to conclude that, the
prospect of observing $\eta_b$ at LHC through the 4$\mu$ mode looks
promising. However, we should be fully aware that it will be a
highly challenging task for experimentalists to single out the
relatively few signal events from the presumedly abundant continuum
background.

To summarize, in this work we have studied the exclusive decay
process $\eta_b \to J/\psi J/\psi$ to NLO in $\alpha_s$ while at LO
in $v_c$. We found that this new contribution to the amplitude is
comparable in magnitude with the previously calculated tree-level
relativistic correction piece, but differs by a phase about
$90^\circ$. Including this new piece of contribution will
substantially enhance the previous estimate of $\mathcal{B}(\eta_b
\to J/\psi J/\psi)$, thus the observation prospect of this clean
$\eta_b$ hadronic decay mode in the forthcoming LHC experiment
becomes brighter.

\acknowledgments

This work is supported in part by National Natural Science
Foundation of China under Grants No.~10605031, 10475083, and by the
Chinese Academy of Sciences under Project No. KJCX3-SYW-N2.

\appendix


\end{document}